\documentclass{ptptex}
\usepackage{wrapft}
\usepackage{graphicx}

\newcommand{\braket}[1]{\langle{#1}\rangle}
\def\beq{\begin{eqnarray}}
\def\eeq{\end{eqnarray}}
\def\bsub{\begin{subequations}}
\def\esub{\end{subequations}}
\def\beq{\begin{eqnarray}}
\def\eeq{\end{eqnarray}}
\def\bsub{\begin{subequations}}
\def\esub{\end{subequations}}
\def\b{\begin{equation}}
\def\bs{\begin{split}}
\def\es{\end{split}}
\def\e{\end{equation}}

\renewcommand{\theequation}{\arabic{section}.\arabic{equation}}

\begin{document}

\title{Spin polarizations under a pseudovector interaction between quarks 
with the Kobayashi-Maskawa-'t Hooft term\\
in high density quark matter}

\author{Masatoshi {\sc Morimoto}$^1$, Yasuhiko {\sc Tsue}$^{2}$, {Jo\~ao da {\sc Provid\^encia}}$^{3}$,\\
{Constan\c{c}a {\sc Provid\^encia}}$^{3}$ 
and {Masatoshi {\sc Yamamura}}$^{4}$
}

\inst{$^{1}${Graduate School of Integrated Arts and Science, Kochi University, Kochi 780-8520, Japan}\\
$^{2}${Department of Mathematics and Physics, Kochi University, Kochi 780-8520, Japan}\\
$^{3}${CFisUC, Departamento de F\'{i}sica, Universidade de Coimbra, 3004-516 Coimbra, Portugal}\\
$^{4}${Department of Pure and Applied Physics, 
Faculty of Engineering Science,\\
Kansai University, Suita 564-8680, Japan}\\
}

\abst{
A possibility of a quark spin polarization originated from a pseudovector condensate 
is investigated in the three-flavor Nambu-Jona-Lasinio model with the 
Kobayashi-Maskawa-'t Hooft interaction which leads to flavor mixing. 
It is shown that a pseudovector condensate related to the strange quark 
easily occurs compared with pseudovector condensate related to light quarks.
Further, it is shown that 
the pseudovector condensate related to the strange quark 
appears at a slightly small chemical potential 
by the effect of the flavor mixing due to the Kobayashi-Maskawa-'t Hooft interaction. 
}


\maketitle

\section{Introduction}

One of recent interests in many-particle systems governed by 
quantum chromodynamics (QCD) is to clarify the existence of 
various phases in high density and finite or zero temperature quark matter.\cite{FH} 
Especially, in quark matter at finite density and low temperature, 
there may exist various phases such as the color superconducting phase 
\cite{ARW,IB,CFL}, the quarkyonic phase,\cite{McL} 
the inhomogeneous chiral condensed phase,\cite{NT}
the quark ferromagnetic phase,\cite{Tatsumi} the color-ferromagnetic phase,\cite{Iwazaki}
the spin polarized phase due to the axial vector interaction\cite{NMT,TMN} 
or due to the tensor interaction
\cite{BJ,IJMP,oursPTP,oursPTEP1,oursPTEP2,oursPTEP3,oursPTEP4,oursPR,Ferrer,MT,MNYY} 
and so forth. 
In order to investigate the phase structure in quark matter, various effective models 
of QCD are used because in the region with low temperature and high density, 
namely large quark chemical potential, the numerical simulation by using the lattice QCD did not work until now, 
while in the region of high temperature and zero density, 
the lattice QCD simulation gives useful information about the phase structure.

As one of the effective models of QCD, the Nambu-Jona-Lasinio (NJL) model \cite{NJL} 
is widely used \cite{Klevansky,HK}
because the NJL model has an important chiral symmetry of QCD. 
This model is used to investigate quark matter in the region with 
large quark chemical potential at low temperature.\cite{Buballa}
Thus, the physics related to the chiral symmetry or chiral symmetry breaking 
is well described. 
By using the extended NJL model in which the tensor-type four-point interaction 
and/or the vector-pseudovector-type four-point interaction between quarks is 
introduced retaining the chiral symmetry, 
the possibility of the tensor condensate and/or psuedovector condensate 
has been investigated.\cite{BJ,IJMP,oursPTP,oursPTEP1,oursPTEP2,oursPTEP3,Ferrer,MT,MNYY,Maedan} 
Since it has been shown that the quark spin polarization leads to the spontaneous 
magnetization in quark matter in the case 
of tensor-type \cite{oursPTEP3} or pseudovector-type interaction \cite{Morimoto} 
between quarks in the NJL model, there exists a possibility that is gives origin to the 
strong magnetic field of compact stars such as neutron stars and magnetars.
\cite{HL}

In this paper, we concentrate on the spin polarization due to the pseudovector 
condensate originated from the pseudovector interaction between quarks in an 
extended NJL model. 
In our previous paper in \citen{Morimoto}, the spin polarization 
due to the pseudovector condensate has been investigated in the case of the 
two-flavor NJL model. 
Then, it has been shown that the pseudovector condensate appears 
in a rather narrow region of the quark chemical potential just before the chiral symmetry is restored. 
In this region, the dynamical quark mass is still not zero. 
However, if quark mass becomes zero, the pseudovector condensate disappears 
even if the strength of pseudovector interaction is very large. 
Thus, there may be a possibility of the existence of 
the pseudovector condensate related to the strange quark because 
the strange quark has a finite current quark mass even in the region with large quark chemical 
potential, namely high density. 
Because the pseudovector condensate leads to spin polarization and spontaneous magnetization, 
it is interesting to investigate a possibility of the pseudovector condensate 
in the three-flavor NJL model. 
In three-flavor case, it is well known that the quark-flavor mixing occurs through 
the six-point interaction between quarks in the NJL model. 
This interaction is called the Kobayashi-Maskawa-'t Hooft interaction or 
the determinant interaction.
\cite{KM,determinant} 
Thus, in this paper, the effect of the flavor mixing on the appearance of 
pseudovector condensate is also investigated.

This paper is organized as follows: 
In the next section, the mean field approximation for the NJL model 
with vector-pseudovector-type four-point interaction between quarks 
is given. Then, both the quark and antiquark condensate, namely chiral condensate, 
and the pseudovector condensate, namely spin polarization, are introduced 
and in section 3, the thermodynamic potential is evaluated at zero temperature 
with finite quark chemical potential. 
Both the condensates are treated self-consistently 
by means of the gap equations. 
In section 4, the solutions of the gap equations are numerically given and 
the behaviors of the pseudovector condensates and the dynamical quark masses 
related to the light quarks ($u$ and $d$ quarks) and the strange quark 
are investigated. 
The last section is devoted to a summary and concluding remarks.

\section{Mean field approximation for the Nambu-Jona-Lasinio model with vector-pseudovector-type four-point interaction between quarks}

Let us start from the three-flavor Nambu-Jona-Lasinio model with vector-pseudovector-type \cite{NMT,TMN} 
four-point interactions between quarks. 
The Lagrangian density can be expressed as 
\beq\label{2-1}
& &{\cal L}=\mathcal{L}_0+\mathcal{L}_m+\mathcal{L}_S+\mathcal{L}_{P}+\mathcal{L}_{D}, 
\nonumber\\
& &{\cal L}_0={\bar \psi}i\gamma^\mu \partial_{\mu}\psi , \nonumber \\
& &{\cal L}_m=-{\bar \psi} \vec m_0\psi , \nonumber \\
& &{\cal L}_S=\frac{G_s}{2}\sum^8_{a=0}[({\bar \psi}\lambda_a\psi)^2+({\bar \psi}i\lambda_a\gamma_5\psi)^2],\nonumber \\
& &{\cal L}_P=-\frac{G_p}{2}\sum^8_{a=0}[({\bar \psi}\gamma^\mu\lambda_a\psi)^2+({\bar \psi}i\gamma_5\gamma^\mu\lambda_a\psi)^2] , \nonumber \\
& &{\cal L}_D=G_D\left[\text{det}\bar\psi(1-\gamma_5)\psi + \text{det}\bar\psi(1+\gamma_5)\psi\right],
\eeq
where $\vec m_0$ represents a current quark mass matrix in flavor space 
as follows : 
\begin{align}\label{2-2}
	\vec m_0&= \text{diag}\left( m_u,m_d,m_s \right) \ .
\end{align}
Here, ${\cal L}_P$ represents a four-point vector and pseudovector interaction between quarks in the three-flavor case 
which preserves chiral symmetry. 
Also, ${\cal L}_D$ represents so-called the Kobayashi-Maskawa-'t Hooft or the determinant 
interaction term
which leads to the six-point interaction between quarks in the three-flavor case.

Hereafter, we treat the above model within the mean field approximation. 
First, we ignore non-diagonal components of the condensates in a flavor space. 
Therefore, terms in a summation of Gell-Mann matrices are restricted to the diagonal entries 
with $a=0, 3$ and 8 : 
\begin{align}\label{2-3}
	\sum^{8}_{a=0}[({\bar \psi}\lambda_a\Gamma\psi)^2] &
\longrightarrow
 \sum_{a=0,3,8}[({\bar \psi}\lambda_a\Gamma\psi)^2]\nonumber \\
	&\quad
=\frac{2}{3}\left[\left(\bar u \Gamma u +\bar d \Gamma d +\bar s\Gamma s\right)\right]^2 
	+\left[ \left(\bar u\Gamma u - \bar d\Gamma d\right)\right]^2 \nonumber\\
	& \qquad +\frac{1}{3}\left[\left(\bar u\Gamma u +\bar d\Gamma d -2\bar s\Gamma s\right)\right]^2 \nonumber \\
	&\quad
= 2(\bar u\Gamma u)^2 +2(\bar d\Gamma d)^2+2(\bar s\Gamma s)^2 \ .
\end{align}
Here, $\Gamma$ means products of any gamma matrices or unit matrix.
Also, in the determinant interaction term, ${\cal L}_D$, 
the same approximation is adopted, namely, the off-diagonal matrix elements in 
the flavor space are omitted: 
\begin{align}\label{2-4}
	&\text{det}\bar\psi\left(1-\gamma_5\right)\psi + \text{det}\bar\psi\left(1-\gamma_5\right)\psi \nonumber \\ 
	&\longrightarrow
 \text{det}
	\begin{pmatrix}
		\bar u(1-\gamma_5)u & 0 & 0 \\
		 0 & \bar d(1-\gamma_5)d &0 \\
		0 & 0 &\bar s(1-\gamma_5)s 
	\end{pmatrix}
\nonumber\\
&\qquad + \text{det}
	\begin{pmatrix}
		\bar u(1+\gamma_5)u & 0 & 0 \\
		0 & \bar d(1+\gamma_5)d &0 \\
		0 & 0 &\bar s(1+\gamma_5)s 
	\end{pmatrix} \nonumber\\
	&= 2(\bar uu) (\bar dd) (\bar ss) 
\nonumber\\
&\quad
+ 2(\bar uu) (\bar d\gamma_5d) (\bar s\gamma_5s)+2(\bar u\gamma_5u) (\bar dd) (\bar s\gamma_5s)+2(\bar u\gamma_5u) (\bar d\gamma_5d) (\bar ss) \ .
\end{align}
Secondly, in order to consider the spin polarization under the mean field approximation, 
the pseudovector condensate $\braket{{\bar q}\gamma_5\gamma^3 q}=\braket{{q^\dagger}\Sigma_3 q}$
is taken into account as well as the chiral condensate $\langle {\bar q}q\rangle$. 
It should be noted that the pseudovector condensate for  
$\braket{{\bar q}\gamma_5\gamma^\nu q}$ with $\nu=3$ 
is nothing but the expectation value 
of the spin matrix $\Sigma_3$ for the quark number density $q^{\dagger}q$. 
Thus,  the pseudovector condensate 
$\braket{{\bar q}\gamma_5\gamma^3 q}$ can be regarded as a quark spin polarization.
Then, the Lagrangian density (\ref{2-1}) reduces to 
\begin{align}
\label{2-5}
\mathcal L_{MF}=&{\bar \psi}(i\gamma^\mu \partial_\mu -\vec M_q)\psi-\vec U{\psi^{\dagger}}\Sigma_3\psi \nonumber \\
&-\sum_f\left(\frac{M(f)^2}{4G_s}+\frac{U_f^2}{4G_p}\right) +\frac{1}{2}\frac{G_D}{G_s^3}M(u)M(d)M(s)\ , 
\end{align}
where $f=u,\ d$ or $s$ and 
\begin{align}
&\Sigma_3=-\gamma^0\gamma_5\gamma^3=
\left(
\begin{array}{cc}
\sigma_3 & 0 \\
0 & \sigma_3
\end{array}
\right)
\nonumber\\
&\vec M_q =\text{diag.}\left(m_u + M(u) -\frac{G_D}{2G_s^2}M(d)M(s)\ ,\ m_d + M(d) -\frac{G_D}{2G_s^2}M(s)M(u)\ , \right.
\nonumber\\
&\qquad\qquad\qquad \left. \ m_s + M(s) -\frac{G_D}{2G_s^2}M(u)M(d) \right) \nonumber\\
&\qquad =\text{diag.}(M_u,\ M_d,\ M_s) \nonumber\\
	&M(f) = -2G_s\braket{\bar q_f q_f} \label{2-6}\\
	&\vec U=\text{diag.}\left(U_u\ ,\ U_d\ ,\ U_s\right) \nonumber\\
	&U_f = -2G_p\braket{q_f^\dagger \Sigma_3 q_f} . \label{2-7}
\end{align}
Here, $\sigma_3$ is the third component of the Pauli spin matrices.

Introducing the quark chemical potential $\mu$ in order to consider a quark matter 
at finite density, 
the Hamiltonian density can be obtained from the Lagrangian density as
\beq\label{2-8}
{\cal H}_{MF}-\mu{\cal N}
&=&{\bar \psi}\left(-i{\mib \gamma}\cdot {\mib \nabla}+\vec M_q-\mu\gamma^0+\vec U\gamma^0\Sigma_3
\right)\psi
\nonumber\\
& &
+\sum_f\left(\frac{M(f)^2}{4G_s}+\frac{U_f^2}{4G_p}\right) -\frac{1}{2}\frac{G_D}{G_s^3}M(u)M(d)M(s)\ , 
\eeq
where ${\cal N}$ represents the quark number density, $\psi^{\dagger}\psi$.

\section{Thermodynamic potential}

In this section, let us derive the effective potential or the thermodynamic potential at zero temperature. 
The Hamiltonian density (\ref{2-8}) 
can be rewritten as 
\beq
{\cal H}_{MF}-\mu{\cal N}
&=&\psi^{\dagger}(h_A-\mu)\psi +\sum_f\left(\frac{M(f)^2}{4G_s}+\frac{U_f^2}{4G_p}\right) -\frac{1}{2}\frac{G_D}{G_s^3}M(u)M(d)M(s),
\ \  \label{3-9}\\
h_A&=&-i \gamma^0{\mib\gamma}\cdot {\mib \nabla}+ \gamma^0\vec M_q+ \vec U\Sigma_3 .\label{3-10} 
\eeq
In order to obtain the eigenvalues of the single-particle Hamiltonian $h_A$, namely the energy eigenvalues of single quark, it is necessary to diagonalize $h_A$, 
the eigenvalues of which can be obtained easily as 
\begin{align}
\label{3-11}
	E^f_{p_x,p_y,p_z,\eta} &=
\sqrt{p_x^2+p_y^2+\left(\sqrt{p_z^2+M_{f}^2}+\eta U_f\right)^2}\ ,
\end{align} 
where $\eta=\pm 1$. 
Thus, we can easily evaluate the thermodynamic potential with the above single-particle energy eigenvalues. 
Then, the thermodynamic potential $\Phi$ can be expressed as
\begin{align}
\label{3-12}
	\Phi =
	&\sum_{f,\alpha,\eta} \int \frac{dp_z}{2\pi}\int \frac{dp_x}{2\pi}\int \frac{dp_y}{2\pi}  \left(E^f_{p_x,p_y,p_z,\eta}-\mu\right) \theta\left(\mu-E^f_{p_x,p_y,p_z,\eta}\right) \nonumber \\
		&-\sum_{f,\alpha,\eta} \int \frac{dp_z}{2\pi}\int \frac{dp_x}{2\pi}\int \frac{dp_y}{2\pi} E^f_{p_x,p_y,p_z,\eta} \nonumber \\
		&+ \sum_f\left(\frac{M(f)^2}{4G_s} + \frac{U_f^2}{4G_p}\right) + \frac{1}{2}\frac{G_D}{G_s^3}M(u)M(d)M(s).
\end{align}
Here, $\theta(x)$ represents the Heaviside step function. 
The first and second lines in (\ref{3-12}) represent the positive-energy contribution of quarks and the vacuum contribution, respectively. 
To make the calculation easier, we substitute $p_x^2+p_y^2 \equiv p_\bot^2$. 
Then, the energy eigenvalues and the thermodynamic potential are rewritten as
\begin{align}
	E^f_{p_x,p_y,p_z,\eta} &=\sqrt{p_\bot^2+\left(\sqrt{p_z^2+M_{qf}^2}+\eta U_f\right)^2}\ \nonumber \\
	& \equiv E^f_{p_\bot,p_z,\eta} \ ,\label{3-13}\\
	\Phi 
		 =&\sum_{f,\alpha,\eta} \int \frac{dp_z}{2\pi}\int \frac{dp_\bot}{2\pi}  p_\bot\left( E^f_{p_\bot,p_z,\eta}-\mu\right) \theta\left(\mu-E^f_{p_\bot,p_z,\eta}\right) \nonumber \\
		&-\sum_{f,\alpha,\eta} \int \frac{dp_z}{2\pi} \int \frac{dp_\bot}{2\pi} p_\bot E^f_{p_\bot,p_z,\eta} \nonumber \\
		&+ \sum_f\left(\frac{M(f)^2}{4G_s} + \frac{U_f^2}{4G_p}\right) + \frac{1}{2}\frac{G_D}{G_s^3}M(u)M(d)M(s) .
\label{3-14}
\end{align}


Noting the condition $\left( \mu>E^f_{p_\bot,p_z,\eta}\right )$ due to the step function, 
integration ranges of positive-energy contribution should be carefully estimated.
First of all,
\begin{align}
&0 \le p_\bot \le \sqrt{\mu^2-\left(\sqrt{p_z^2+M_q^2}+\eta U\right)^2}
\label{eq:pbot}
\end{align}
should be satisfied. 
Next, for the $\eta=+1$ case, integration range of $p_z$ is
\begin{align}
\label{eq:sekibunhanni+1}
	|p_z| \le \sqrt{(\mu-U)^2-M_q^2} \ 
\end{align}
because $p_\bot$ is real. 
On the other hand, for $\eta=-1$ case, the followings are obtained: 
\begin{align}
\label{eq:sekibunhanni-1}
\left \{
\begin{array}{ll}
 {\rm for}\ \ U \le M_q \ \cdots\ |p_z| \le \sqrt{(\mu+U)^2-M_q^2}  \\
{\rm for}\ \ M_q \le U  \ \cdots \ 
\left\{
\begin{array}{ll}
{\rm for}\ \ \mu \le U  \ \cdots \  \sqrt{(U-\mu)^2-M_q^2} \le |p_z| \le \sqrt{(U+\mu)^2-M_q^2} \\
{\rm for}\ \ U \le \mu  \ \cdots \ |p_z| \le \sqrt{(U+\mu)^2-M_q^2} \ .
\end{array} \right . 
\end{array} \right . 
\end{align}

As for the vacuum contributions, since the NJL model is not a renormalizable model, 
the three-momentum cutoff $\Lambda$ is usually introduced as 
\beq\label{3-18}
p_x^2+p_y^2+p_z^2 \leq \Lambda^2\ . 
\eeq
Thus, the thermodynamic potential (\ref{3-12}) can be divided into tow parts and can be evaluated as follows : 
\begin{align}
\label{eq:Phitasu}
\Phi&=
\Phi_{\mu}+\Phi_{vac}\ ,
\end{align}
\begin{align}
\label{eq:Phi_mu}
\Phi_{\mu}=\frac{3}{2\pi^2} \sum_f \Biggl[ &\frac{p_z}{24}\left(12U\mu -12U^2 -5M_q-2p_z^2\right)\sqrt{M_q^2+p_z^2} + 4p_z^3\left(\mu-2U\right)   \nonumber \\
&- p_z\left(24M_q^2U + 12M_q^2\mu - 8U^3 + 12U^2\mu -4\mu^3\right) 
 \nonumber \\
&-\frac{1}{8}M_q^2\left(M_q^2+4U^2-4U\mu\right) \text{ln} \left(p_z+\sqrt{M_q^2+p_z^2}\right)\ \Biggl]^{p_{Max}}_{0} 
\nonumber \\
+\frac{3}{2\pi^2} \sum_f &\Biggl[ -\frac{p_z}{24}\left(12U\mu +12U^2 +5M_q+2p_z^2\right)\sqrt{M_q^2+p_z^2} + 4p_z^3\left(\mu+2U\right)   \nonumber \\
&\quad - p_z\left(-24M_q^2U + 12M_q^2\mu + 8U^3 + 12U^2\mu -4\mu^3\right)
 \nonumber \\
&\quad -\frac{1}{8}M_q^2\left(M_q^2+4U^2+4U\mu\right) \text{ln} \left(p_z+\sqrt{M_q^2+p_z^2}\right)\ \Biggl]^{p_{Max}}_{p_{min}} 
 \ ,
\end{align}
\begin{align}
\label{eq:Phi_vac}
\Phi_{vac}=\frac{1}{8\pi^2}\sum_f&
\left[\Lambda\sqrt{\Lambda^2+M_q^2}(5M_q^2+2\Lambda^2+12U^2)+3M_q^2(M_q^2+4U^2)\ln\frac{\Lambda+\sqrt{\Lambda^2+M_q^2}}{M_q}\right]
\nonumber\\
&-\frac{1}{2\pi^2}\sum_f\int_0^{\Lambda}dp_z
\Biggl[\left(\Lambda^2-p_z^2+\left(\sqrt{p_z^2+M_q^2}-U\right)^2\right)^{\frac{3}{2}}
\nonumber\\
&\qquad\qquad\qquad\qquad\qquad
+\left(\Lambda^2-p_z^2+\left(\sqrt{p_z^2+M_q^2}+U\right)^2\right)^{\frac{3}{2}}\Biggl]\nonumber\\
&+ \sum_f\left(\frac{M(f)^2}{4G_s} + \frac{U_f^2}{4G_p}\right) + \frac{1}{2}\frac{G_D}{G_s^3}M(u)M(d)M(s) \ .
\end{align}
Here, $\Phi_{vac}$ represents the contribution of vacuum. 
In Eq.(\ref{eq:Phi_mu}),  $[f(x)]^a_b$ means definite integral, namely $f(a)-f(b)$. 
Here, we defined $p_{Max}$ and $p_{min}$, based on (\ref{eq:sekibunhanni+1}) and (\ref{eq:sekibunhanni-1}), explicitly,
\begin{align}\label{3-22}
	&p_{Max} \equiv \sqrt{(\mu+U)^2-M_q^2} \quad \text{or} \quad 0 \nonumber \\
	&p_{min} \equiv \sqrt{(U-\mu)^2-M_q^2} \quad \text{or} \quad 0 \ .
\end{align}
Thus, the thermodynamic potential can be calculated analytically, expect for the second term of  (\ref{eq:Phi_vac}).

\section{Numerical results}

In this section, we give numerical results. 
Especially, the effect of the determinant interaction on the pseudovector condensate 
is considered. 

First, we switch off the pseudovector interaction, $G_P=0$, namely, the 
pseudovector condensate being zero, $U=0$. 
Then, the thermodynamic potential $\Phi$, which represents $\Phi_{U=0}$ in this 
case, is written as
\begin{align}
\label{4-23}
\Phi_{U=0}=&\frac{3}{8\pi^2}\sum_f\left[\frac{1}{3}\sqrt{\mu^2-M_q^2}(-2\mu^3+5\mu M_q^2)-M_q^4\ln\frac{\mu+\sqrt{\mu^2-M_q^2}}{M_q}\right]
\theta(\mu-M_q)\nonumber\\
& -\frac{3}{8\pi^2}\sum_f\left[\Lambda\sqrt{\Lambda^2+M_q^2}(2\Lambda^2+M_q^2)-M_q^4\ln\frac{\Lambda+\sqrt{\Lambda^2+M_q^2}}{M_q}\right]
\nonumber\\
& + \sum_f\frac{M(f)^2}{4G_S}  + \frac{1}{2}\frac{G_D}{G_S^3}M(u)M(d)M(s) \ . 
\end{align} 
To determine the chiral condensates or the constituent quark masses, 
the gap equation is derived as 
\begin{align}
\label{4-24}
\frac{\partial \Phi_{U=0}}{\partial M(u)} = \frac{\partial \Phi_{U=0}}{\partial M(d)} = \frac{\partial \Phi_{U=0}}{\partial M(s)} = 0\ .
\end{align}
From here, we assume isospin symmetry, namely $m_u=m_d$ and 
$\langle{\bar u}u\rangle=\langle{\bar d}d\rangle$. 
If we adopt model parameters written in Table \ref{table:modelparameters}., the dynamical quark masses $M_{u,d}\equiv M_q$ and $M_s$ are obtained as $M_q=0.335$ GeV and $M_s=0.527$ GeV, respectively.
If we neglect the determinant interaction, namely $G_D=0$, then 
we have to adopt $G_S=4.370/\Lambda^2$ instead of $3.666/\Lambda^2$ in 
Table \ref{table:modelparametersU}. 
It should be noted that the dynamical quark masses $M_q$ and $M_s$ do not 
depend on parameter $G_P$. 
Thus, it is allowed that $G_P$ can be regarded as a free parameter in this model.  


\begin{table}[b]
\caption{Parameter set of 3-flavor NJL model.}
\label{table:modelparameters}
\begin{center}
\begin{tabular}{c|c|c|c|c}
\hline
$\Lambda$ & $G_S$ & $G_D$ & $m_{u,d}$ & $m_s$\\
0.6314 GeV &$ 3.666/\Lambda^2$ & $-9.288/\Lambda^5$&0.0055 GeV &0.1357 GeV \\
\hline
\end{tabular}
\end{center}
\end{table}


\begin{table}[t]
\caption{Parameter sets.}
\label{table:modelparametersU}
\begin{center}
\begin{tabular}{c|cccc}
\hline
Model & $\Lambda$ & $G_S$ & $G_D$ &$G_P$\\
\hline\hline
GP0 & 0.6314GeV &$ 3.666/\Lambda^2$ & $-9.288/\Lambda^5$&$0$ \\
\hline
GP2 & 0.6314GeV &$ 3.666/\Lambda^2$ & $-9.288/\Lambda^5$&$2G_S$ \\
\hline
GP4.1 & 0.6314GeV &$ 3.666/\Lambda^2$ & $-9.288/\Lambda^5$&$4.1G_S$ \\
\hline
GP5 & 0.6314GeV &$ 3.666/\Lambda^2$ & $-9.288/\Lambda^5$&$5G_S$ \\
\hline\hline
GP0GD0 & 0.6314GeV &$ 4.370/\Lambda^2$ & $0$&$0$ \\
\hline
GP2GD0 & 0.6314GeV &$ 4.370/\Lambda^2$ & $0$&$2 \times 3.666/\Lambda^2$ \\
\hline
GP4.1GD0 & 0.6314GeV &$ 4.370/\Lambda^2$ & $0$&$4.1 \times 3.666/\Lambda^2$ \\
\hline
GP5GD0 & 0.6314GeV &$ 4.370/\Lambda^2$ & $0$&$5 \times 3.666/\Lambda^2$ \\
\hline
\end{tabular}
\end{center}
%
\end{table}

Let us assume $U\geq 0$ without loss of generality.
%
As is well known, the chiral condensate appears in the vacuum due to the vacuum contribution. 
However, the psuedovector condensate does not appear in the vacuum in a realistic case.
Thus, in the following parts of this section, we neglect the vacuum contribution to the pseudovector condensate 
for simplicity of the numerical calculation.
We concentrate our attention on the appearance of the pseudovector condensate 
due to the particle-contribution in the quark matter only.
%
We adopt model parameter sets in Table \ref{table:modelparametersU}.
In these parameters, the pseudovector interaction strength $G_P$ is taken as a free 
parameter in our consideration because this parameter could not be determined by the experimental 
values of a certain physical quantity. 
Therefore, by using these parameter sets, we investigate behavior of the pseudovector condensate in finite density quark matter.
Further, to investigate the effects of the determinant interaction term $\mathcal{L_D}$, 
we include parameter sets, namely model GPnGD0 without the determinant interaction, 
$G_D=0$.  
In these parameter sets, the dynamical quark masses $M_q$ and $M_s$ 
have almost the same values as those of the case $G_D \neq 0$.

\begin{figure}[b]
    \begin{tabular}{cc}
	\begin{minipage}[t]{0.45\hsize}
	\begin{center}
		\includegraphics[height=4cm]{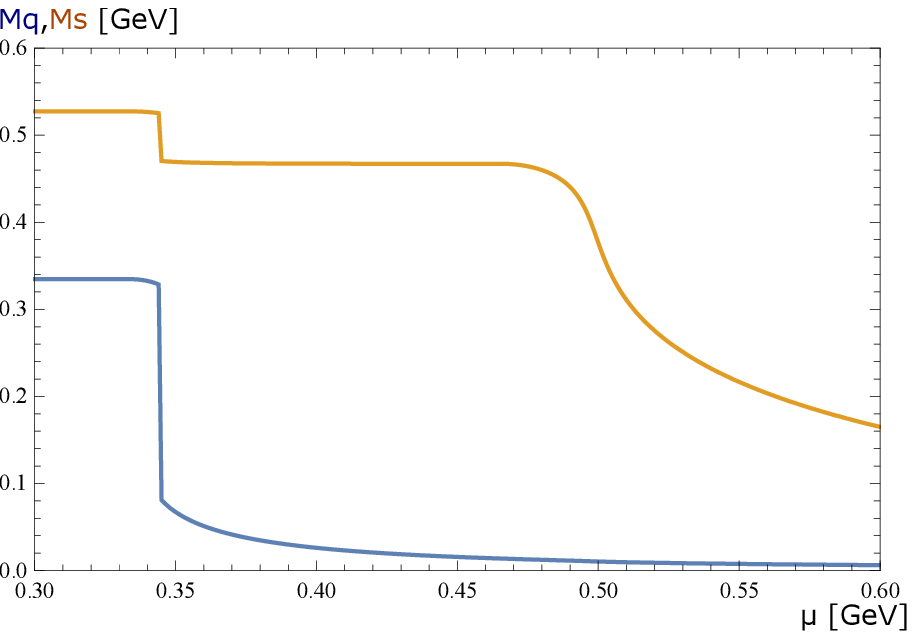}
	\caption{Quark masses $M_q$ (lower curve) and $M_s$ (upper curve) 
are depicted as a function of chemical potential $\mu$ in model GP0.
}
\label{fig:GP0}
	\end{center}
	\end{minipage}
\qquad
	\begin{minipage}[t]{0.45\hsize}
	\begin{center}
		\includegraphics[height=4cm]{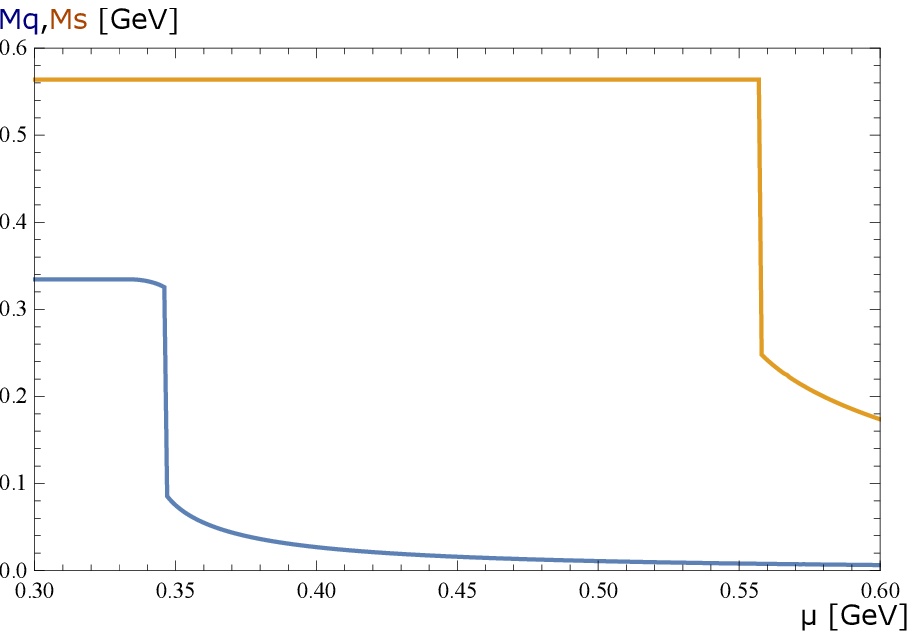}
	\caption{Quark masses $M_q$ (lower curve) and $M_s$ (upper curve) 
are depicted as a function of chemical potential $\mu$ in model GP0GD0.
}
\label{fig:GP0GD0}
	\end{center}
	\end{minipage}
   \end{tabular}
\end{figure}

In the following we investigate the effect of $G_P$. 
We will consider three different values of $G_P$: $2G_S$, $4.1G_S$ and $5G_S$, 
which correspond to only the appearance of the strange quark pseudovector condensate, 
the onset of the light quark pseudovector condensate besides the strange quark condensate, and existence of 
both light quark and $s$-quark condensates at different chemical potentials.
In model GP0, namely the original 3-flavor NJL model with the determinant interaction, 
the chiral symmetry is broken and the non-trivial solution of the gap equation for chiral condensate or dynamical quark mass exists in $\mu<\mu_{\rm cr1}= 0.34$ GeV for 
light quarks. 
However, in $\mu > \mu_{\rm cr1}$, the chiral symmetry is restored and $M_{q}$ has only a small value due to the current quark mass. 
For strange quark, in $\mu = \mu_{\rm cr1}$, the value of dynamical quark mass $M_s$ decreases by the effect of the chiral restored light quarks.
Also, in $\mu > \mu_{\rm cr2} \approx 0.527$ GeV, 
$M_{s}$ decreases monotonically. 
These behavior is plotted Fig.\ref{fig:GP0}.
On the other hand, as is seen in Fig.\ref{fig:GP0GD0} 
for model GP0GD0 without the determinant interaction, 
the dynamical quark mass $M_q$ for light quarks and $M_{s}$ for the strange quark 
are independent each other because there is no flavor mixing caused by the determinant interaction.
These results show that strange quark mass $M_{s}$ is strongly affected by the effect of flavor mixing term.

\begin{figure}[t]
    \begin{tabular}{cc}
	\begin{minipage}[t]{0.45\hsize}
	\begin{center}
		\includegraphics[height=4cm]{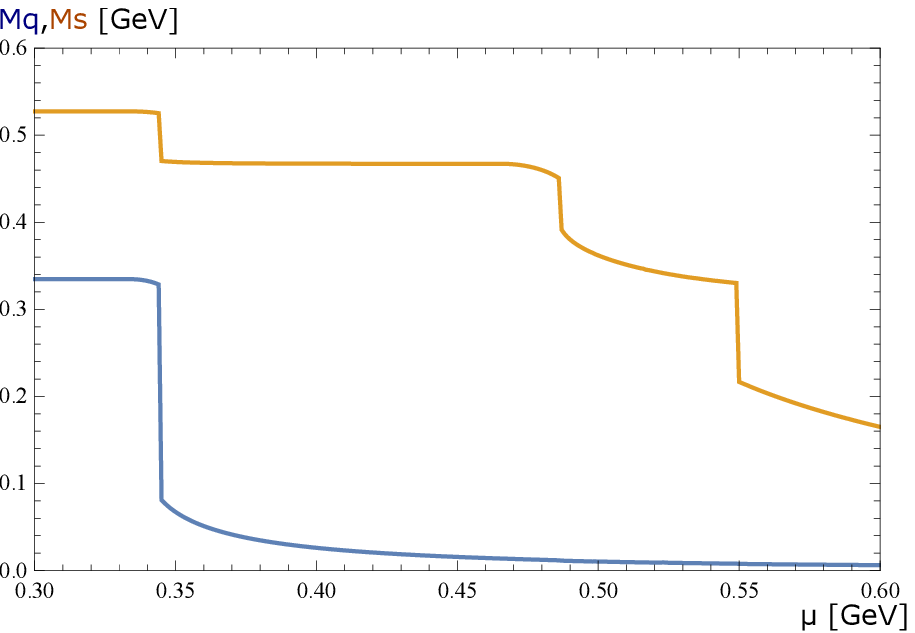}
	\caption{Quark masses $M_q$ (lower curve) and $M_{s}$ (upper curve) 
are depicted as a function of chemical potential $\mu$ in model GP2.
}
\label{fig:GP2M}
	\end{center}
	\end{minipage}
\qquad
	\begin{minipage}[t]{0.45\hsize}
	\begin{center}
		\includegraphics[height=4cm]{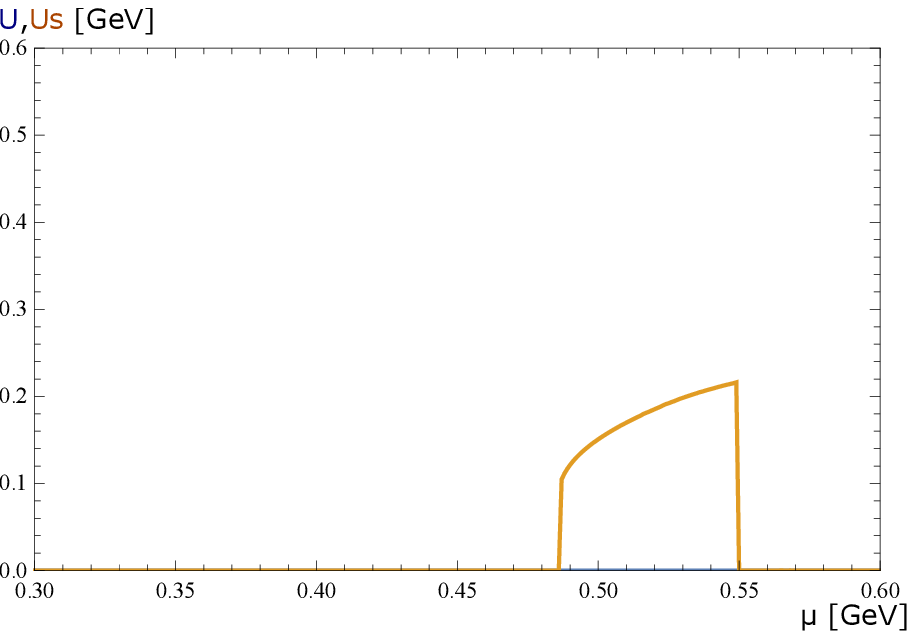}
	\caption{Pseudovector condensate $U_s$ is depicted as a function of chemical potential $\mu$ in model GP2. 
}
\label{fig:GP2U}
	\end{center}
	\end{minipage}
   \end{tabular}
\end{figure}

\begin{figure}[t]
    \begin{tabular}{cc}
	\begin{minipage}[t]{0.45\hsize}
	\begin{center}
		\includegraphics[height=4cm]{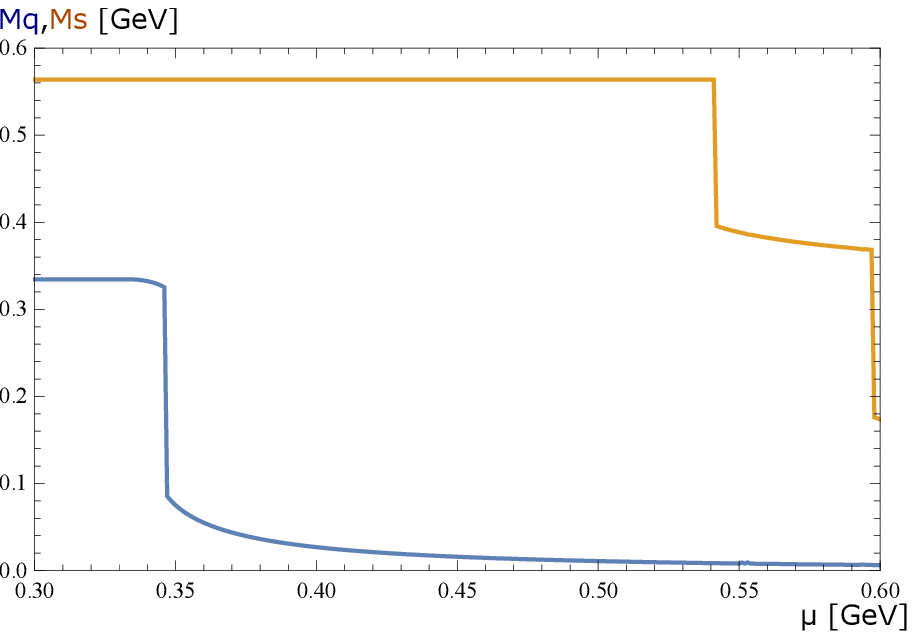}
	\caption{Quark masses $M_q$ (lower curve) and $M_{s}$ (upper curve) 
are depicted as a function of chemical potential $\mu$ in model GP2GD0.
}
\label{fig:GP2GD0M}
	\end{center}
	\end{minipage}
\qquad
	\begin{minipage}[t]{0.45\hsize}
	\begin{center}
		\includegraphics[height=4cm]{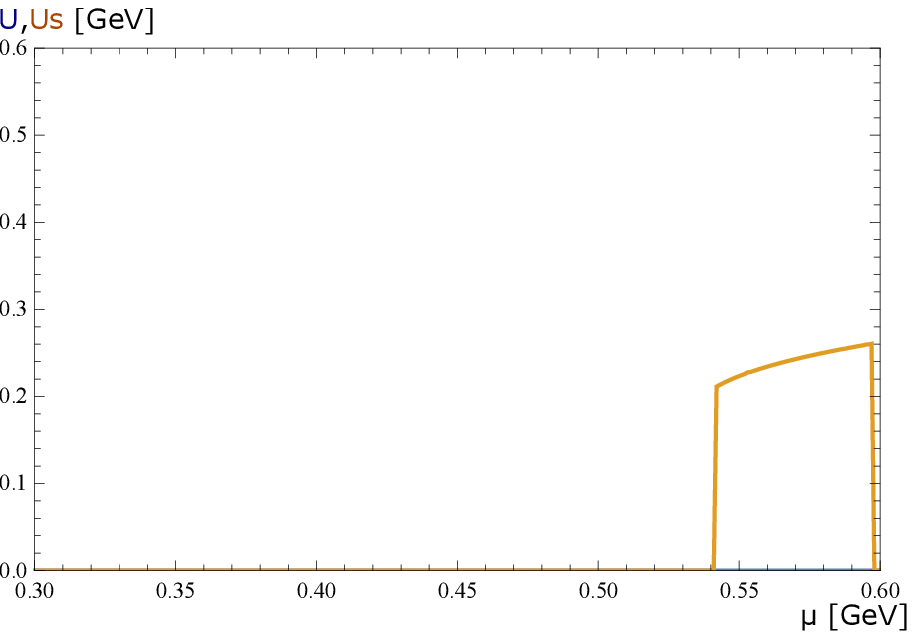}
	\caption{Pseudovector condensate $U_s$ is depicted as a function of chemical potential $\mu$ in model GP2GD0. 
}
\label{fig:GP2GD0U}
	\end{center}
	\end{minipage}
   \end{tabular}
\end{figure}

\begin{figure}[t]
    \begin{tabular}{cc}
	\begin{minipage}[t]{0.45\hsize}
	\begin{center}
		\includegraphics[height=4cm]{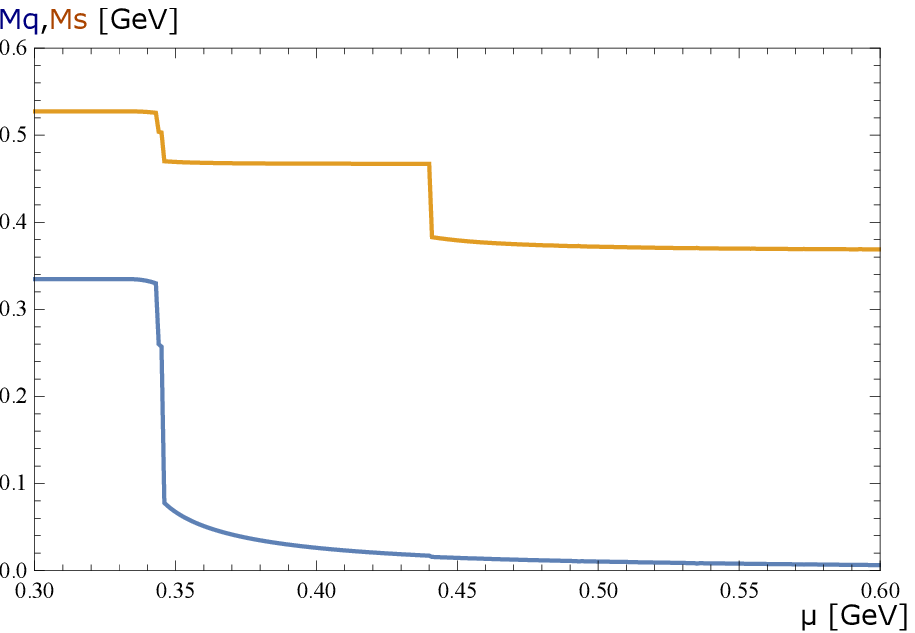}
	\caption{Quark masses $M_q$ (lower curve) and $M_{s}$ (upper curve) 
are depicted as a function of chemical potential $\mu$ in model GP4.1.
}
\label{fig:GP4.1M}
	\end{center}
	\end{minipage}
\qquad
	\begin{minipage}[t]{0.45\hsize}
	\begin{center}
		\includegraphics[height=4cm]{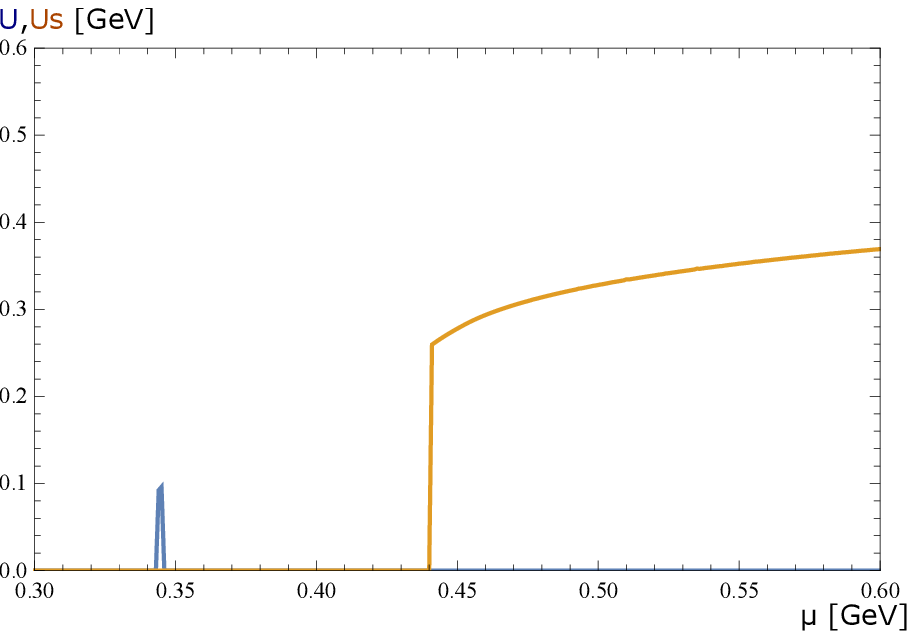}
	\caption{Pseudovector condensates $U_q$ (left)  and $U_s$ (right) 
are depicted as a function of chemical potential $\mu$ in model GP4.1. 
}
\label{fig:GP4.1U}
	\end{center}
	\end{minipage}
   \end{tabular}
\end{figure}

\begin{figure}[t]
    \begin{tabular}{cc}
	\begin{minipage}[t]{0.45\hsize}
	\begin{center}
		\includegraphics[height=4cm]{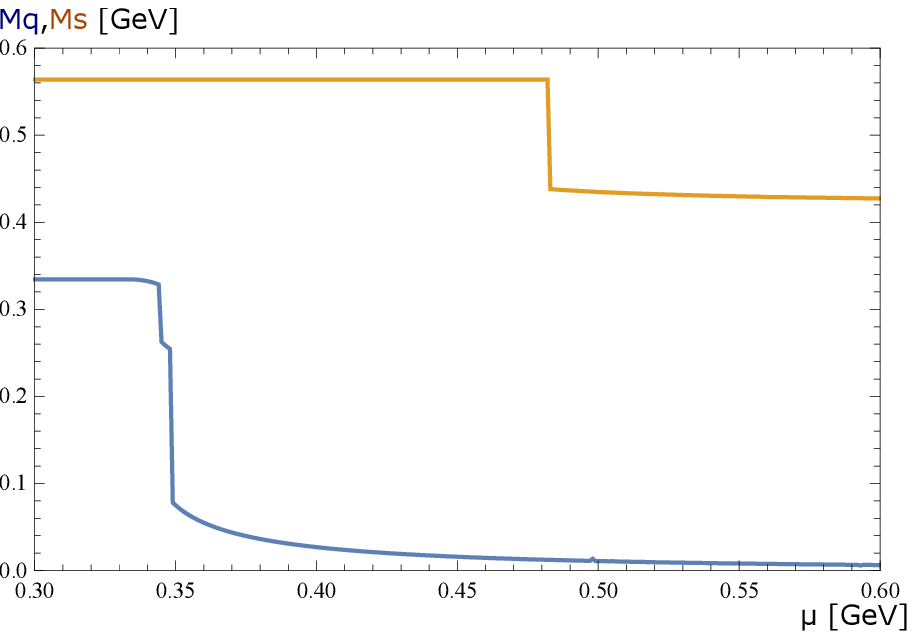}
	\caption{Quark masses $M_q$ (lower curve) and $M_{s}$ (upper curve) 
are depicted as a function of chemical potential $\mu$ in model GP4.1GD0.
}
\label{fig:GP4.1GD0M}
	\end{center}
	\end{minipage}
\qquad
	\begin{minipage}[t]{0.45\hsize}
	\begin{center}
		\includegraphics[height=4cm]{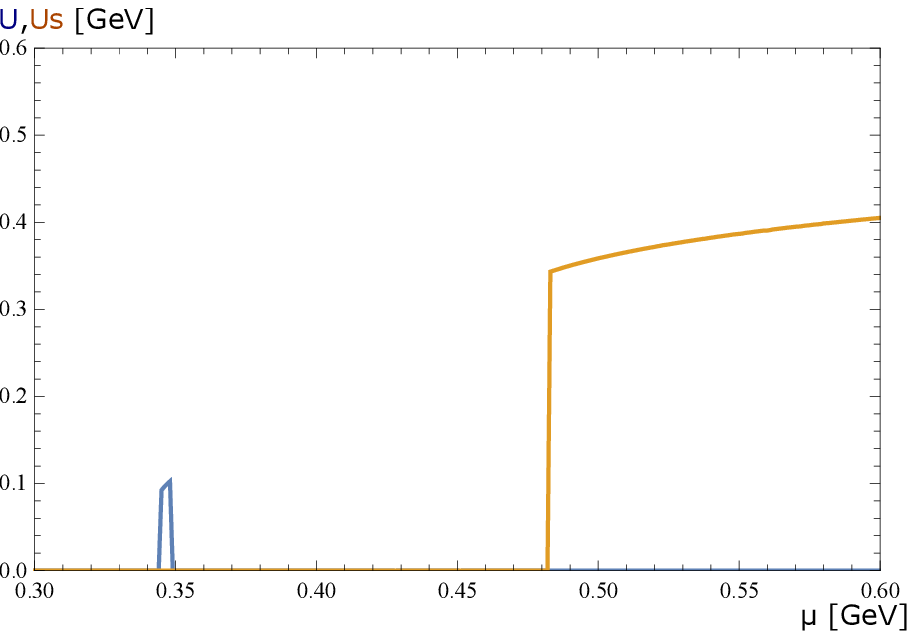}
	\caption{Pseudovector condensates $U_q$ (left) and $U_s$ (right) 
are depicted as a function of chemical potential $\mu$ in model GP4.1GD0. 
}
\label{fig:GP4.1GD0U}
	\end{center}
	\end{minipage}
   \end{tabular}
\end{figure}

In models with $G_P\neq0$, pseudovector condensates $U_{u,d}\equiv U_q$ and $U_s$ appear.
In Figs.\ref{fig:GP2M} and \ref{fig:GP2U},
 the dynamical quark masses and the pseudovector condensate $U_s$ are depicted as a function of the quark chemical potential $\mu$ in model GP2. 
As is seen in Fig.\ref{fig:GP2M}, the dynamical quark mass $M_q$ for light quarks 
jumps at $\mu_{\rm cr1}$ and for $\mu>\mu_{\rm cr1}$ the dynamical quark mass monotonically decreases because the chiral symmetry is restored. 
On the other hand, for the dynamical quark mass of strange quark, 
first, at $\mu=\mu_{\rm cr1}$, the mass jumps slightly. 
Secondly, at $\mu\approx \mu_{{\rm cr}U_s1}\approx 0.486$ GeV, 
the dynamical quark mass decreases again. 
This behavior is originated from the appearance of the pseudovector condensate 
$U_s$ due to strange quark. 
Finally, at $\mu \geq \mu_{{\rm cr}U_s2}\approx 0.55$ GeV, 
the dynamical quark mass jumps again and the pseudovector condensate 
$U_s$ disappears.
%
If the pseudovector condensates do not appear,
the dynamical quark masses decrease smoothly as is seen in Fig.1 due to 
the chiral symmetry restoration. 
However, when the pseudovector condensates appear, the decrease of the dynamical quark 
masses is suppressed. 
Thus, the fine structure of the behavior of dynamical quark mass as a function 
of the quark chemical potential, as is seen in Fig.3, appears.
%

In Figs.\ref{fig:GP2GD0M} and \ref{fig:GP2GD0U}, 
the dynamical quark masses and the pseudovector condensate $U_s$ are shown 
as a function of $\mu$ in model GP2GD0 without the determinant interaction. 
The light quark mass and the strange quark mass show a rather simple behavior. 
Also, the pseudovector condensate $U_s$ appears in a certain range of $\mu$. 
This behavior is similar to the one of the pseudovector condensate $U_q$ 
instead of $U_s$ in the two-flavor NJL model shown in our previous paper. 
Further, it is shown that 
the critical chemical potential $\mu_{{\rm cr}U_s1}$ at which the pseudovector 
condensate  $U_s$ for the strange quark appears is smaller than the one without 
the determinant interaction, namely, $\mu_{{\rm cr}U_s1}\approx 0.486 / 0.541$ GeV 
in the case with/without the determinant interaction. 
In models GP2 and GP2GD0, only the pseudovector condensate $U_s$ appears. 
Under this model parameter $G_P=2G_S$, $U_q$ does not appear.

Figures \ref{fig:GP4.1M} and \ref{fig:GP4.1U} are the same as Figs.\ref{fig:GP2M} and \ref{fig:GP2U} 
except for the value of $G_P$, namely $G_P=4.1G_S$. 
%
As is seen in Fig.\ref{fig:GP4.1M}, both the dynamical quark masses $M_q$ and $M_s$ 
jump slightly at $\mu_{{\rm cr}U_q1} \approx 0.342$ GeV.
Then, at $\mu=\mu_{{\rm cr}U_q2} \approx 0.343$ GeV, the dynamical quark masses jump again. 
In the narrow region $\mu_{{\rm cr}U_q1}<\mu<\mu_{{\rm cr}U_q2}$, the pseudovector condensate for the light quarks 
$U_q$ appears as is seen in Fig.\ref{fig:GP4.1U}.
For $\mu>\mu_{{\rm cr}U_q2}$, the dynamical quark mass decreases monotonically. 
On the other hand, the strange quark has the finite dynamical mass about 0.48 GeV. 
At $\mu=\mu_{{\rm cr}U_s1}\approx 0.441$ GeV, the dynamical quark mass jumps again. 
Simultaneously, at $\mu=\mu_{{\rm cr}U_s1}$, the pseudovector condensate 
$U_s$ sets in, see Fig.\ref{fig:GP4.1U}.

\begin{figure}[t]
    \begin{tabular}{cc}
	\begin{minipage}[t]{0.45\hsize}
	\begin{center}
		\includegraphics[height=4cm]{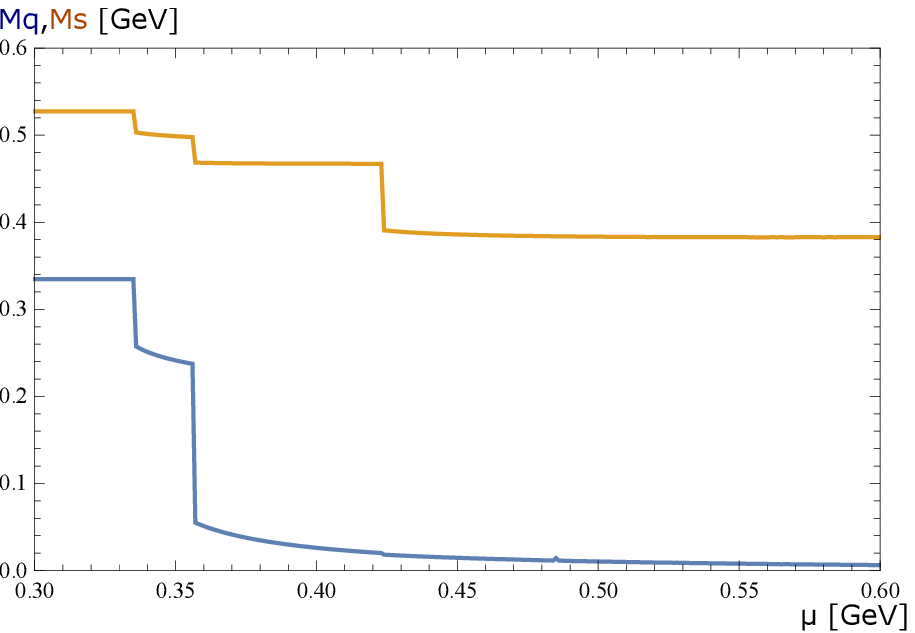}
	\caption{Quark masses $M_q$ (lower curve) and $M_{s}$ (upper curve) 
are depicted as a function of chemical potential $\mu$ in model GP5.
}
\label{fig:GP5M}
	\end{center}
	\end{minipage}
\qquad
	\begin{minipage}[t]{0.45\hsize}
	\begin{center}
		\includegraphics[height=4cm]{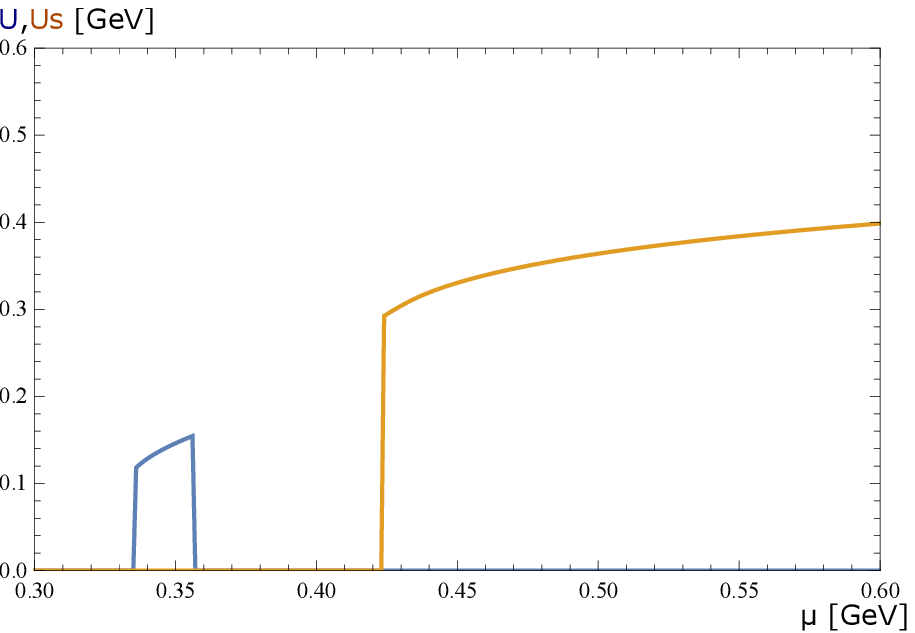}
	\caption{Pseudovector condensates $U_q$ (left) and $U_s$ (right) 
are depicted as a function of chemical potential $\mu$ in model GP5. 
}
\label{fig:GP5U}
	\end{center}
	\end{minipage}
   \end{tabular}
\end{figure}
\begin{figure}[t]
    \begin{tabular}{cc}
	\begin{minipage}[t]{0.45\hsize}
	\begin{center}
		\includegraphics[height=4cm]{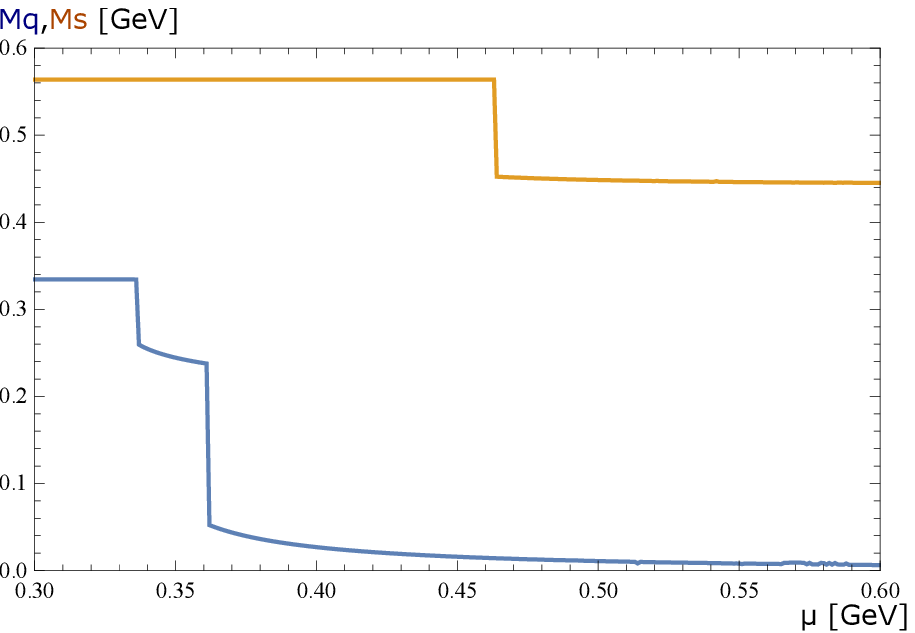}
	\caption{Quark masses $M_q$ (lower curve) and $M_{s}$ (upper curve) 
are depicted as a function of chemical potential $\mu$ in model GP5GD0.
}
\label{fig:GP5GD0M}
	\end{center}
	\end{minipage}
\qquad
	\begin{minipage}[t]{0.45\hsize}
	\begin{center}
		\includegraphics[height=4cm]{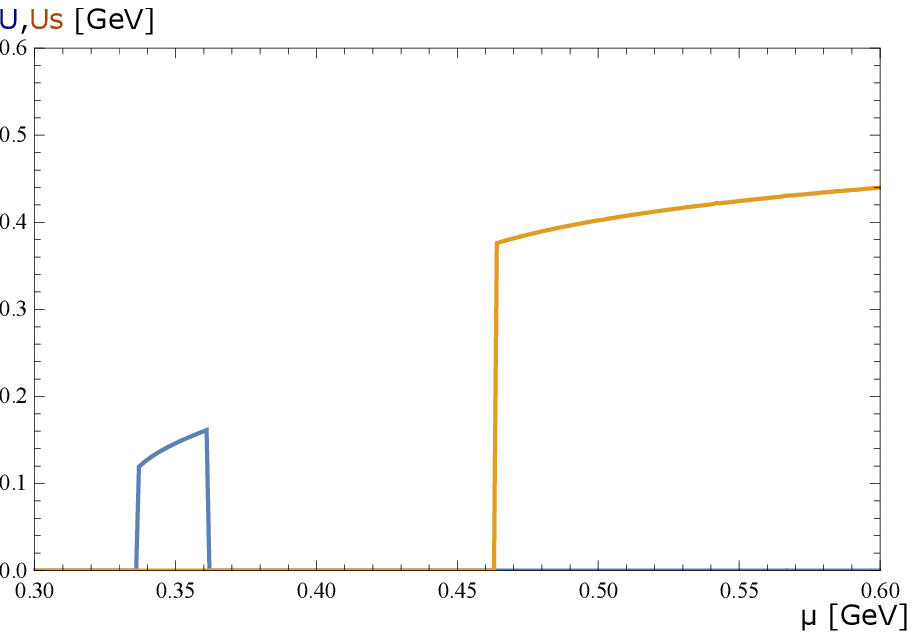}
	\caption{Pseudovector condensate $U_q$ (left) and $U_s$ (right) 
are depicted as a function of chemical potential $\mu$ in model GP5GD0. 
}
\label{fig:GP5GD0U}
	\end{center}
	\end{minipage}
   \end{tabular}
\end{figure}

Figures \ref{fig:GP4.1GD0M} and \ref{fig:GP4.1GD0U} are the same as Figs.\ref{fig:GP2GD0M} and \ref{fig:GP2GD0U} 
except for the value of $G_P$. 
The behavior of the light quark mass is similar to the behavior of model GP4.1. 
However, the behavior of the strange quark mass is different because 
there is no flavor mixing. 
Therefore, the dynamical quark mass for strange quark is not affected by the 
pseudovector condensate $U_q$ for light quark.  
Also, the pseudovector condensate for the light quarks is not almost affected 
by the determinant interaction. 
As in model GP2GD0, the critical chemical potential for the strange quark $\mu_{{\rm cr}U_s1}$ is smaller than the one without 
the determinant interaction, namely, $\mu_{{\rm cr}U_s1}\approx 0.441/ 0.482$ GeV 
in the case with/without the determinant interaction.

Figures \ref{fig:GP5M} and \ref{fig:GP5U} are the same as Figs.\ref{fig:GP4.1M} and \ref{fig:GP4.1U} except for $G_P=5G_S$. 
This model has a large coupling constant of the pseudovector interaction between quarks. 
In this model, the critical chemical potentials have the values of $\mu_{{\rm cr}U_q1} \approx 0.333$ GeV, $\mu_{{\rm cr}U_q2}\approx 0.358$ GeV and $\mu_{{\rm cr}U_s1}\approx 0.423$ GeV.
The regions in which $U_q$ and $U_s$ exist 
are expanded, compared to model GP4.1.
In this strong coupling case with $G_P=5G_S$, 
the pseudovector condensate does not disappear in $\mu<\Lambda$. 

Figures \ref{fig:GP5GD0M} and \ref{fig:GP5GD0U} are the same as Figs.\ref{fig:GP4.1GD0M} and \ref{fig:GP4.1GD0U} except for value of $G_P$. 
Comparing Figs.\ref{fig:GP5U} with \ref{fig:GP5GD0U}, it is seen that the pseudovector condensate for the light quarks is not almost affected 
by the determinant interaction. 
In the models of GP5 and GP5GD0, the critical chemical potential for the strange quark $\mu_{{\rm cr}U_s1}$ takes the value of
 $\mu_{{\rm cr}U_s1}\approx 0.423/ 0.463$ GeV 
in the case with/without the determinant interaction.

\section{Summary and concluding remarks}

It has been shown that the pseudovector condensate, which leads to 
the quark-spin polarization as was shown in our previous paper \cite{Morimoto}, 
occurs due to the pseudovector-type four-point interaction between quarks in quark matter at zero temperature within the three-flavor NJL model. 
Focusing on the determinant interaction in three-flavor NJL model which leads to the 
quark-flavor mixing, 
we have investigated the effect of flavor mixing  on the dynamical quark masses and the  
pseudovector condensates.
As a result, the quantities related to the strange quark are 
affected by the determinant interaction, especially the behavior of the 
dynamical quark mass as a function of the quark chemical potential, 
while the quantities related to the light quarks are hardly affected. 
The pseudovector condensate related to the strange quark occurs at a rather small 
quark chemical potential compared with the case of no flavor mixing, namely, the case 
without the determinant interaction. 
The different behavior of the quark masses, which depend strongly on the presence of the 
determinant interaction, is the cause of this result.

Under the model parameters used in this paper, 
the pseudovector condensate for light quarks 
and one for the strange quark do not coexist.
It may be necessary to investigate the possibility of the coexistence of both 
the pseudovector condensates due to the light quarks and the strange quark. 
This is one of future problems to solve. 
Also, we have not  explicitly calculated magnetic properties, such that spontaneous magnetization, magnetic susceptibility and so on.  
These are interesting future problems which are left in order to clarify the magnetic properties of high density quark matter.  
Further, the implication to the compact stars such as the 
neutron star and magnetar should be investigated
by assuming the existence of the 
pseudovector condensate related to the light quarks and the strange quark. 
%
In this investigation, it is necessary to impose the beta equilibrium and the charge 
neutrality conditions. 
In the case with tensor-type condensate under the tensor-interaction 
in the NJL model \cite{Matsuoka2}, the influence of the condensate for the hybrid star has been 
investigated under the beta equilibrium and charge neutrality conditions. 
In this paper, it has been also shown that 
the quark chemical potential in which the tensor condensate appears is slightly large compared with no 
beta equilibrium and charge neutrality conditions and the value of the condensate is smaller.  
It may be interesting to evaluate the equation of state with 
pseudovector condensate and to investigate the effect of it on the structure of compact star.
%
This may be interesting future problem.

\section*{Acknowledgements}

Two of the authors (M.M and Y.T.) would like to express their sincere thanks to 
Dr. E. Nakano for his important suggestion. 

\end{document}